\documentstyle[12pt]{article}

\begin{document}
\baselineskip=24pt \parskip=0pt plus2pt
\textheight=22cm
\begin{center}
\begin{Large}
Phase Separation in a Simple Model with Dynamical Asymmetry
\end{Large}

\vskip1.cm

by

Rajeev Ahluwalia 

\vskip1.cm
Jawaharlal Nehru Centre for Advanced Scientific Research, Jakkur,\\
Bangalore, India\\
and\\
Materials Research Centre \footnote{ 
Address for correspondence\\  email:rajeev@mrc.iisc.ernet.in}, 
Indian Institute of Science\\
Bangalore, 560 012, India

\end{center}

\vskip2cm

\begin{abstract}
We perform computer simulations of a Cahn-Hilliard model of phase separation
which has dynamical asymmetry between the two coexisting phases. 
The dynamical asymmetry is incorporated by
considering a mobility function which is order parameter dependent. 
Simulations of this model reveal morphological features similar to those
observed in viscoelastic phase separation. In the early stages, the minority 
phase domains form a percolating 
structure which shrinks with time eventually leading to the formation of 
disconnected domains. The domains grow as $L(t) \sim t^{1/3}$ in the 
very late stages. Although dynamical scaling is violated in the area 
shrinking regime, it is restored at late times. However, the form of the 
scaling function is found to depend on the extent of dynamical asymmetry.  

\end{abstract}

\pagebreak
\section*{I. Introduction}
Phase separation phenomena in binary mixtures have been the subject of much recent 
research in condensed matter physics \cite{jd}. In a typical phase separation experiment, 
a binary mixture ( such as an alloy, polymer blend or a binary liquid mixture) 
is quenched from it's one phase region into a region of it's phase diagram where the
constituent phases tend to segregate. The subsequent dynamics consists of the formation 
and growth of domains which are rich in either of the phases. It is now well 
established that for  
mixtures whose constituent phases have identical dynamical
properties, the domain growth satisfies the dynamical scaling hypothesis 
in the late stages \cite{stau}. 
According to this hypothesis, the equal time structure factor of the appropriate 
order parameter satisfies the scaling law
\begin{equation}
\label{scal1}
S(\vec{k},t)=L(t)^{d}
F(kL(t)),
\end{equation} 
where $F$ is a scaling function and $L(t)$ is a time 
dependent length scale which can 
be associated with the mean size of the growing domains ( $d$ refers to the 
spatial dimension). The dynamical scaling implies   
that the evolution of domains is self-similar, i.e. domain size  
grows but the overall morphology does not change with time. The other interesting aspect 
is
the functional form of the length scale $L(t)$. It is now conclusively established 
that   
for pure and isotropic systems, $L(t) \sim t^{\phi}$, where $\phi$ is the growth exponent 
which crucially depends on the nature of the dynamics. For example in the case of a 
binary alloy where 
 there is no intrinsic dynamical asymmetry between the two phases, the growth is driven by 
 surface tension and is characterized by an exponent $\phi=1/3$ \cite{jd}. This is 
commonly refered 
to as the Lifshitz-Slyozov law which also describes domain growth in polymer solutions and 
blends for shallow quenches. For binary liquids the hydrodynamic interactions are 
important and for this case, the growth exponent $\phi=1$ \cite{hyd}.

Recently, there have been some experiments investigating the role of dynamical asymmetry
between the constituent phases of the phase separating system. The dynamical asymmetry 
usually arises when the characteristic relaxation times of the molecules of the coexisting
phases are different. Tanaka has studied phase separation in deeply quenched       
semi-dilute polymer 
solutions \cite{tanaka1} where the asymmetry arises due to the viscoelastiy of the 
polymer rich domains. In another interesting experiment, Tanaka \cite{tanaka2}has 
investigated domain growth in a polymer blend which is quenched to a temperature which is 
close to the glass transition temperature of the minority species. The common feature of  
these systems is that the time scales of molecular motion of the minority phase 
are much 
slower relative to the other phase. This leads to unusual phase separation which is now
commonly refered to as viscoelastic phase separation.

The main features of viscoelastic phase separation are as follows.  
After an initial incubation regime during which no macroscopic phase separation 
occurs, domains of the more mobile majority phase nucleate and start growing.
 The growth of these domains eventually results in the 
formation of a   
 thin sponge-like percolating network of the minority phase 
(this is in contrast to usual phase 
separation where the minority phase forms isolated droplets). The growth of the majority 
phase domains also leads to an overall shrinking in the volume of the minority 
phase regions. 
The shrinking continues 
until the network breaks up into isolated droplets of the minority phase. 

Taniguchi and Onuki \cite{Onuki} have studied this problem by simulating a viscoelastic 
model which incorporates the coupling between stress and diffusion \cite{Doi} 
for a semi-dilute 
polymer solution. They were able to observe a sponge-like 
network of the minority phase in their simulations. However, they were not able to see 
phase inversion (the eventual 
breaking up of the 
network into isolated minority phase domains) 
within the time scales of their simulations. Subsequently, Tanaka and Araki 
\cite{tanaka3} simulated a 
viscoelastic model with the effects of bulk stress included. Using this model, they were 
able to demonstrate most of the experimentally observed features like the formation of the 
minority phase network which eventually breaks down leading to phase inversion.
 
Although, the viscoelastic models are crucial to explain the experimental observations of
Tanaka, dynamical asymmetry can also be studied in framework of the usual Cahn-Hilliard   
theory of phase separation by making use of an order parameter dependent mobility. 
Sappelt and Jackle \cite{Jackle} have studied domain growth in a system 
where one of the phases freezes into a glassy state. They have considered
 an order parameter dependent mobility which is asymmetric about a fixed 
concentration. 
In their simulations, they found an 
unusual growth mechanism for concentrations where the less mobile glassy 
phase is the majority
phase. However, they did not find a sponge-like structure of the glassy phase 
for the case with low volume fraction of the glassy component.

In this paper, we study dynamically asymmetric phase separation within the framework of
Cahn-Hilliard theory by choosing an appropriate mobility function. We propose a model with 
an order parameter dependent mobility which can model many of the features observed in 
Tanaka's experiment, from the point of view of pattern formation. Unlike 
the viscoelastic theories, we do not incorporate stress fields and the dynamics in our 
model is driven by surface tension only. The effect of dynamical asymmetry 
comes from the 
order parameter dependent mobility function.

The organization of this paper is as follows. In section II, we introduce
our dynamical model. We also explain the modelling of the order parameter dependent 
mobility. In section III, we give 
numerical results for pattern evolution. We also show results for the domain growth law 
and the time-dependent structure factor. Section IV is devoted to a discussion of the 
results and the limitations of the model.
\section*{II. Dynamical Model}
The theory is formulated in terms of an order parameter which  
is the concentration difference between the two species. Since the 
concentration difference is a conserved quantity, the time evolution of a 
scaled 
dimensionless order parameter $\phi(\vec{x},t)$ is described by the equation
\begin{equation}
\label{cahn}
{{\partial \phi(\vec{x},t)}\over{\partial t}}=\vec{\nabla}\cdot 
\left[M(\phi(\vec{x},t))\vec{\nabla}\left(-\phi(\vec{x},t) 
+{\phi(\vec{x},t)}^3-{\nabla}^2{\phi(\vec{x},t)}\right)\right],
\end{equation}
where $\vec{x}$ and $t$ are respectively the scaled space and time variables and      
$M(\phi)$ is the mobility function. This is the deterministic Cahn-Hilliard equation
which is also refered to as the Model B in the Halperin and Hohenberg system of 
classification \cite{Halp}. In conventional theories of spinodal decomposition,
the mobility functiom $M(\phi)$ is usually treated as a constant. However, recently 
there have been some studies where the effect of an order parameter dependent 
mobility on the dynamics of phase separation has been investigated \cite{puri}.

In this paper, we consider a mobility function of the type
\begin{equation}
\label{mob}
M(\phi)={{1}\over {1+exp(\alpha\phi-\beta{\phi}^2)}},
\end{equation}
where $\alpha$ and $\beta$ are positive constants $(\beta > \alpha)$. The 
motivation 
for choosing this particular form of the mobility is as follows. In the 
early stages of domain growth 
$\phi$ is small and for a large enough value of $\alpha$, the mobility is a 
sharp step function 
about $\phi=0$. 
 The negative 
quadratic term on the other hand provides a competing effect on the 
dynamical asymmetry as $\phi$ increases. This term is responsible for 
weakening
of dynamical asymmetry in the late stages and is crucial to get phase 
inversion. The effect of this term on the dynamics is in some sense 
analogous to stress 
relaxation in viscoelastic systems.

\section*{III. Numerical Results}
In this section, we give details of our numerical simulations of phase separation 
for an off-critical quench into the unstable region, using the above described model.  
 We solve equation (\ref{cahn}) with the mobility function given in equation
(\ref{mob}) on an $N \times N $ square lattice with periodic boundary 
conditions. A simple Euler discretization is used with mesh size $\Delta x=1.2$ and the
smallest time step $\Delta t=0.02$.
The initial condition are given
by
\begin{equation}
\label{init}
\phi(\vec{r},0)=\bar{\phi} +\delta\phi(\vec{r},0),
\end{equation}
where $\bar{\phi}$ is the off-criticality and $\delta\phi(\vec{r},0)$ represents 
random fluctuations
uniformly distributed in the interval $[-0.005,0.005]$.
In the simulations reported in this paper, we choose
$\bar{\phi}=-0.1$ which corresponds 
to a minority phase concentration of $0.45$. 

We first describe our results on pattern evolution on an $N \times N$
lattice with $N=128$.
We consider a quench 
corresponding
to  $(\alpha=100,\beta=160)$.  
In figure 1 we display the
evolution of domains corresponding to $\bar{\phi}=-0.1$.
The darker contrast
regions correspond to the minority phase $(\phi > 0)$ and the
brighter regions correspond
to the majority phase $(\phi < 0)$. The shade varies with the extent of order which is characterized
by the local value of the order parameter. In the very early stages,
the growth is strongly influenced by the dynamical asymmetry. At
$t=0$, the system is in a one phase state corresponding to
$\bar{\phi}=-0.1$. As order parameter fluctuations start getting
amplified, the growth of concentration in regions which are
locally rich in the minority component is suppressed due to low
mobility. However, regions which are rich in the majority
component order much faster ( this is in contrast to usual phase
separation where both minority and majority phases order rapidly
and the minority phase forms isolated droplets ). The snapshot
at time $t=50$ in figure 1 corresponds to this situation where
we can see the emergence of local regions rich in the majority
phase. These regions are more ordered as compared to the minority phase
regions. However, the boundaries between the two phases are
still not very sharp ( this is analogous to the so called
incubation regime in viscoelastic phase separation ). When the
order parameter in the majority phase regions reaches it's
saturation value $\phi_{eq}=-1$ ($t \sim 100$), well defined domains of the
majority phase appear and start growing ( keep in mind that the
order parameter in the minority phase regions is yet to reach
it's saturation value of $\phi_{eq} =1$ ). In this regime, the
partially ordered minority phase regions form a percolating
structure whose area keeps on shrinking with time. This thinning
is due to diffusion from the minority phase regions to the
majority phase regions (the minority phase regions tend to expel
the dissolved majority phase component and this results in the growth of
order parameter within the minority phase regions). 

The growth of the majority phase domains and the associated area shrinking can
be clearly seen in the snapshots at times
$t=100$ and $t=200$.    
As the order parameter in the minority phase grows, the negative quadratic term  
in the mobility starts dominating and the dynamics
becomes faster.
 The order parameter in these regions rapidly saturates to the equilibrium 
value ${\phi_{eq}}=1$ 
 (the thick black patches in minority phase  
at time $t=200$ correspond to such regions). 
At this stage, we should also remark that 
the negative quadratic term in the mobility is crucial to observe 
substantial area shrinking and eventual phase inversion. In the absence of this term, the 
mobility of 
the minority phase regions remains 
low for all time, thereby arresting the growth of order parameter\cite{Jackle}.

The area shrinking continues till the order parameter in most of the 
minority phase regions has also
reached its saturation value ${\phi_{eq}}=1$. Notice that by this time, the asymmetry in 
the 
mobility has also disappeared as $M(\phi=1)=M(\phi=-1)$.
 Subsequently, the domain 
growth is expected to occur by the usual Lifshitz-Slyozov or 
evaporation-deposition mechanism, where there is a diffusion from regions 
of higher to lower curvature. Thus domains like to minimise the surface area and
the connectivity of the minority phase regions is expected to break. 
This can be seen from the snapshots at times $t=300$, $t=400$ and 
$t=1000$, where we can see
the appearance of disconnected  minority phase domains. The disconnected 
domains tend 
to relax to circular shapes eventually, as seen in the 
snapshot at time $t=1000$.

We now present results pertaining to dynamical scaling. The quantity of interest here is the time-dependent structure 
factor defined as
\begin{eqnarray}
S(\vec{k},t)={ {\langle \phi(\vec{k},t)\phi(-\vec{k},t) \rangle} \over { {1}\over{N^2}}{\sum_{k}  \langle 
\phi(\vec{k},t)\phi(-\vec{k},t) \rangle}},
\end{eqnarray}
where $\phi(\vec{k},t)$ is the fourier transform of $\phi(\vec{r},t)-\bar{\phi}$ 
and angular brackets refer to an average over initial
conditions. The wavevector k ranges over the first brillouin zone. For the results presented in this paper, we make use of the
isotropy of the system and evaluate a spherically averaged structure factor which depends only on the magnitude of the wavevector.

 We test whether the spherically averaged structure factor obeys the dynamical scaling form
\begin{equation}
S(k,t)=L(t)^dF(kL(t)),
\end{equation}
where $L(t)$ is a length scale related to the mean size of the growing domains. We use the inverse of the first moment of the spherically averaged structure 
factor
as a measure of this length scale, i.e., $L(t) \sim {\langle k \rangle (t) }^{-1}$, where
\begin{equation}
\langle k\rangle(t)={ {{ { \int_0 }^{k_m}} dk k S(k,t) }\over { { { \int_0 }^{k_m} }dk  S(k,t) }}.
\end{equation}
The upper cutoff is taken to be the half the magnitude of the largest wavevector lying in the first brillouin zone.

Before we describe our results on dynamical scaling and the 
structure factor, it is useful to identify the different regimes of growth. 
In figure 2, we plot the area fraction $\phi_A$ of the minority phase regions with the 
dimensionless time variable of our simulations. The quantity $\phi_A$ has been 
obtained by solving equation (2) on an $N\times N$ lattice 
($N=256$)
 and computing the fraction of sites with $\phi > 0$ at each time step. The 
data presented in figure 2 is obtained by averaging over $50$ independent
systems. We see that the area fraction initially increases above it's 
equilibrium value of $0.45$. This corresponds to the fact that the minority 
phase forms
a percolating matrix in the early stages. Subsequently, the area fraction $\phi_A$ rapidly 
decreases. This corresponds to the 
regime in which the concentration within the domains keeps on changing as 
there is a desorption from the minority phase to the majority phase leading 
to area shrinking. 
The area shrinking continues till the
order parameter saturates to it's equilibrium value every where 
$(t \sim 300)$.  
The area fraction saturates close to the equilibrium value of $0.45$ in the late 
stages. 
This regime 
can be clearly seen in
figure 2 for times greater than $t \sim 300$. 
The domain growth in this regime is characterized by the usual 
curvature driven mechanism. We should remark here that very similar time-dependence 
of the volume fraction has been observed in deeply quenched polymer 
blends by Tanaka\cite{tanaka2}.

We now present our results for the structure factor and the length scales. 
We have computed the spherically averaged structure factor and the associated 
 length scale $L(t)$ on a $ 256 \times 256 $ 
lattice by averaging
over $50$ independent initial conditions. In figure 3, we show the 
behaviour of $L(t)$ with $t$ ( $t$ is a dimensionless time variable) 
on a log-log scale.
 We observe an initial fast growth which corresponds to the 
area shrinking regime. The curve crosses over to a straight line which is 
nearly parallel to the solid line of slope $1/3$, thereby indicating that 
our data
conforms to a growth law $L(t) \sim t^{1/3}$ asymptotically. This 
growth law corresponds to the 
regime where both minority and majority phase regions are fully ordered and 
the evaporation-deposition mechanism is expected to apply. 

To test for dynamical scaling hypothesis, we plot $L^{-2}(t) S(k,t)$ vs. $kL(t)$ 
in figure 4, for times $100,200$ and $300$. These times fall within the 
area shrinking regime. We observe that there is no data collapse indicating a 
violation of dynamical scaling for these times. This can be understood if we 
consider the fact that in
the area shrinking regime, the order parameter with in the domains is not 
saturated but keeps on changing with time. However, there is a good data 
collapse at later
times as shown in figure 5. The data at times $600,800$ and $1000$ scales 
well (except in the tail where the finite interfacial width is responsible 
for
deviations from scaling \cite{oono}). In this regime, the order parameter is 
saturated every where and growth takes place by usual 
evaporation-condensation 
mechanism.

It is interesting to compare the form of the scaling function with dynamical 
asymmetry to that with the symmetric mobility case.
In figure 6, we plot $L^{-2}(t)S(k,t)$ with $kL(t)$ at time 1000 for the case of constant 
mobility and the 
dynamically asymmetric case considered in this paper ( the 
data for the symmetric mobility case has been obtained 
for $M(\phi)=1$, with the same initial conditions 
and statistics as the asymmetric mobility case ). We find 
that the two scaling functions have different form. In 
particular, the usual Porod's `shoulder' is less pronounced in the 
dynamically asymmetric case than in the constant mobility case. This 
suggests that the form of the scaled structure factor is dependent on 
the extent of the dynamical asymmetry, for the same value of the 
initial composition.
\section*{IV. Summary and Discussion}
In this paper, we have presented results of computer simulations of a simple 
Cahn-Hilliard type model which has dynamical asymmetry built in through an order parameter
dependent mobility. The form of the mobility function is chosen so as to 
incorporate the effects of a strong dynamical asymmetry in the early 
stages along with a competing term which restores symmetry in the late stages.
Our simple model captures many of the experimentally observed features which have also been observed in simulations on viscoelastic models. Our simulations reveal a 
morphology in which the minority phase forms a percolating structure in the early stages. The area of the minority phase matrix shrinks with time and eventually the matrix starts breaking up into disconnected domains.

We have also tested for the existence of dynamical scaling. We find that the structure factor does not scale very well in the area shrinking regime. 
However, it crosses over into a scaling form when the growth is determined by the evaporation deposition mechanism. 
Interestingly, the form of the scaling function is different than the constant mobility case. 
This suggests that the scaling behaviour is dependent on the extent of the 
dynamical asymmetry ( this can be checked in experiments by considering the dependence of the 
structure factor on the quench depth ). However, the asymptotic growth law is same as that observed in constant mobility systems, i.e., $L(t) \sim t ^{1/3}$.

Although we have been able to account for many of the experimental features, we do not claim this model to be an accurate description of viscoelastic phase separation. We have considered a very simple model which shows growth regimes similar to viscoelastic phase separation. The incorporation of stress fields is essential to
obtain the thin networklike morphologies as observed in experiments, where as in our model, the domain shapes are determined by concentration gradients only. We
should also point out that the percolating minority phase structure is formed in our model only for a small range 
of compositions between $\bar{\phi}=-0.1$ and $\bar{\phi}=0.0$, only for 
a sufficiently large value of the asymmetry parameter $\alpha$.  
Infact for lower $\alpha$, even for the same composition $\bar{\phi} =-0.1$, we do not get an initial percolating
minority phase. The only effect of asymmetry for such cases is on the shape of the domains.

Finally, we remark that our choice of the mobility function is 
not unique. We could construct other forms of the mobility function which could give 
similar results. However, we expect that the associated phase separation to fall into
the same dynamical universality class for all these models. In the present work, we have
attempted to demonstrate that the unusual phase separation observed in viscoelastic 
systems is a more general phenomena, which is expected to show up in systems 
where there is a dynamical asymmetry which gradually decreases as the system approaches
 equlibrium. 
\section*{Acknowledgements}
The author would like to thank Prof. G. Ananthkrishna, Prof.
S. Ramaswamy and Dr. S. Puri for useful discussions. The author would also like to thank 
the 
members of the theory group of Materials Research Center, IISc for their help and 
cooperation.
Financial support from JNCASR
is also gratefully acknowledged.


\newpage
\center{Figure Captions}
\begin{itemize}
\item[Figure 1:] Time evolution of the domains for the asymmetric
mobility case. The dark contrast regions in the snapshots correspond to
the minority phase regions ($\phi > 0$) and the bright contrast regions
correspond to the majority phase ($\phi< 0$). The shade varies with the extent of
ordering determined by the local value of the order parameter. The snapshots 
correspond to times $t=50,100,200,300,400$ and $1000$.
\item[Figure 2:] Variation of the area fraction $\phi_A$ of the minority phase 
with the dimensionless time variable of the simulations.
\item[Figure 3:] Log-log plot of the mean domain size $L(t)$ 
 (Inverse of the first moment of the spherically averaged structure factor)
with the dimensionless time $t$ of the simulations. The solid line 
has a slope $1/3$ and serves as a guide to the eye.   
\item[Figure 4:] Test for dynamical scaling in the early stages. We plot 
$L^{-2}(t)S(k,t)$ vs. $kL(t)$ on a log-log scale for times $t=100,200$ and 
$300$.
\item[Figure 5:] Analogous to figure 4, but for times corresponding to $t=600,
800$ and $1000$.
\item[Figure 6:] Log-log plot of $L^{-2}(t)S(k,t)$ vs $kL(t)$ 
at time $t=1000$ for
the aymmetric mobility case and the constant mobility case with $M(\phi)=1$.  
\end{itemize}  

\newpage         
\begin{thebibliography}{40}
\bibitem{jd} For reviews, see J. D. Gunton, M. San Miguel and P. S. Sahni,
in {\underbar {Phase transitions and Critical Phenomena}} Vol. 8 (ed.
C. Domb and J.L. Lebowitz), p. 267, Academic Press, New York;

K. Binder, in \underbar{Materials Science and Technology,}

\underbar{Vol.5 : Phase Transformations of Materials} (ed.
R. W. Cahn,  P. Haasen and E. J. Kramer), p. 405, VCH, Weinheim (1991);

A. J. Bray, Adv. in Physics \underbar {43}, 357 (1994).

\bibitem{stau} K.Binder and D. Stauffer, Phys. Rev. Lett.
\underbar{33}, 1006 (1974);

also Z. Phys. B \underbar{24}, 406 (1976).

\bibitem{hyd} For numerical simulations, see T. Koga and K. Kawasaki,
Phys. Rev. A \underbar {44}, R817 (1991);

S. Puri and B. Dunweg, Phys. Rev. A \underbar {45}, R6977 (1992);

A. Shinozaki and Y. Oono, Phys. Rev. E \underbar {48}, 2622 (1993).


\bibitem{tanaka1} H. Tanaka, Phys. Rev. Lett. \underbar{71}, 3158 (1993).

\bibitem{tanaka2} H. Tanaka, Phys. Rev. Lett. \underbar{76}, 787 (1996).

\bibitem{Onuki} T. Taniguchi and A. Onuki, Phys. Rev. Lett. 
\underbar{77}, 4910 (1996).

\bibitem{Doi} M. Doi and A. Onuki, J. Phys. II (France) 
\underbar{2}, 1631 (1992)

\bibitem{tanaka3} H. Tanaka and T. Araki, Phys. Rev. Lett. 
\underbar{78}, 4966 (1997).

\bibitem{Jackle} D. Sappelt and J. Jackle, Euro. Phys. Lett. 
\underbar{37}, 13 (1997).

\bibitem{Halp}B. I. Halperin and P. C. Hohenberg, Rev. Mod. Phys. 
\underbar{49}, 435 (1977).

\bibitem{puri} A. M. Lacasta, A. Hernandez-Machado, J. M. Sancho
 and R. Toral, Phys. Rev. B \underbar{45}, 5276 (1992);

A. J. Bray and C. E. Emmott, 
Phys. Rev. B \underbar{52}, R685 (1995);  

S. Puri, A. J. Bray and J. L. Lebowitz, Phys. Rev. E 
\underbar{56}, 758 (1997).
\bibitem{oono} Y. Oono and S. Puri, Mod. Phys. Lett. B \underbar{2}, 861
(1988).

\end{thebibliography}
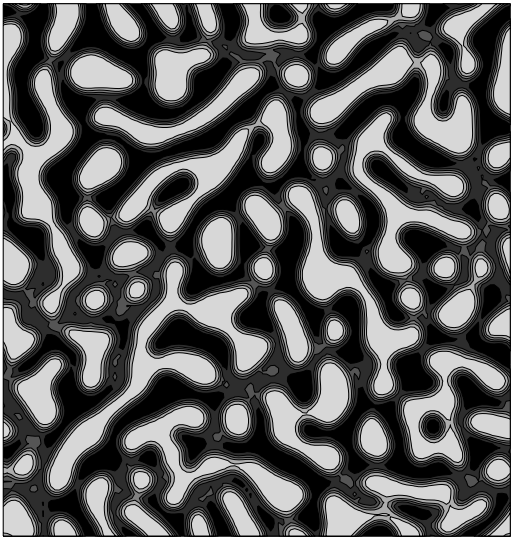
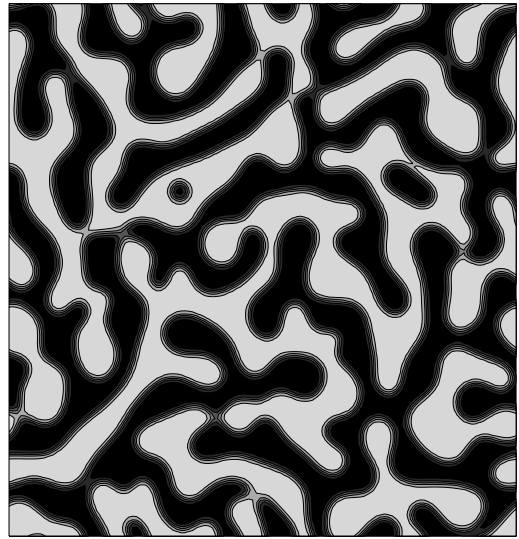
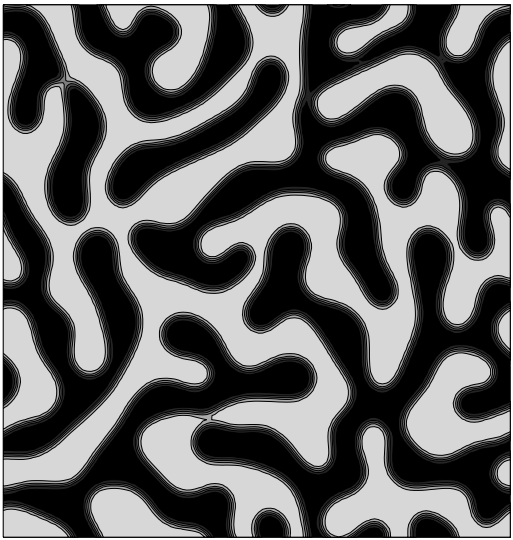
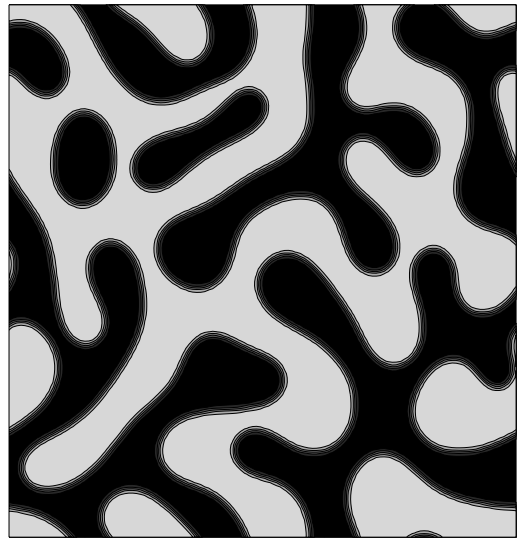
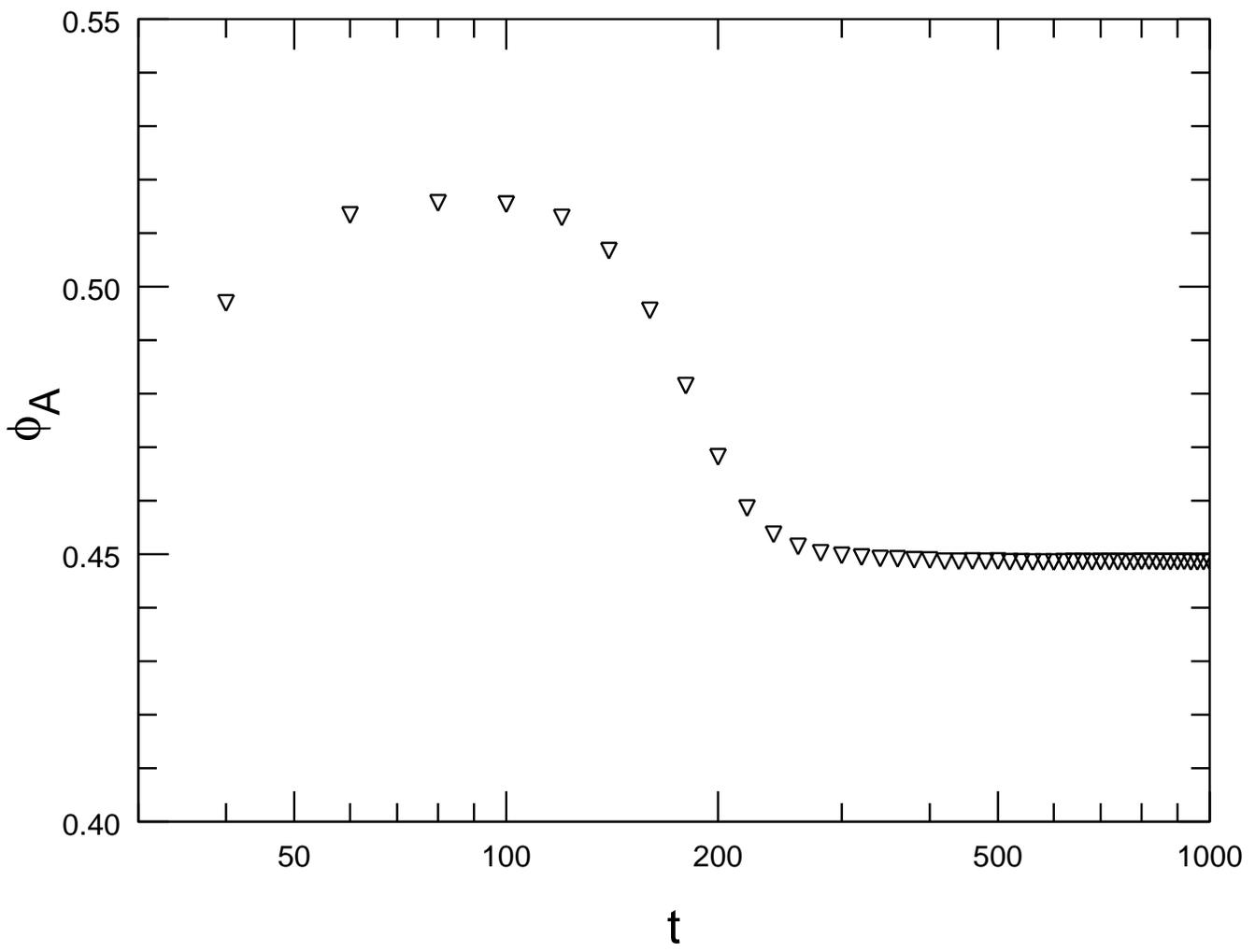
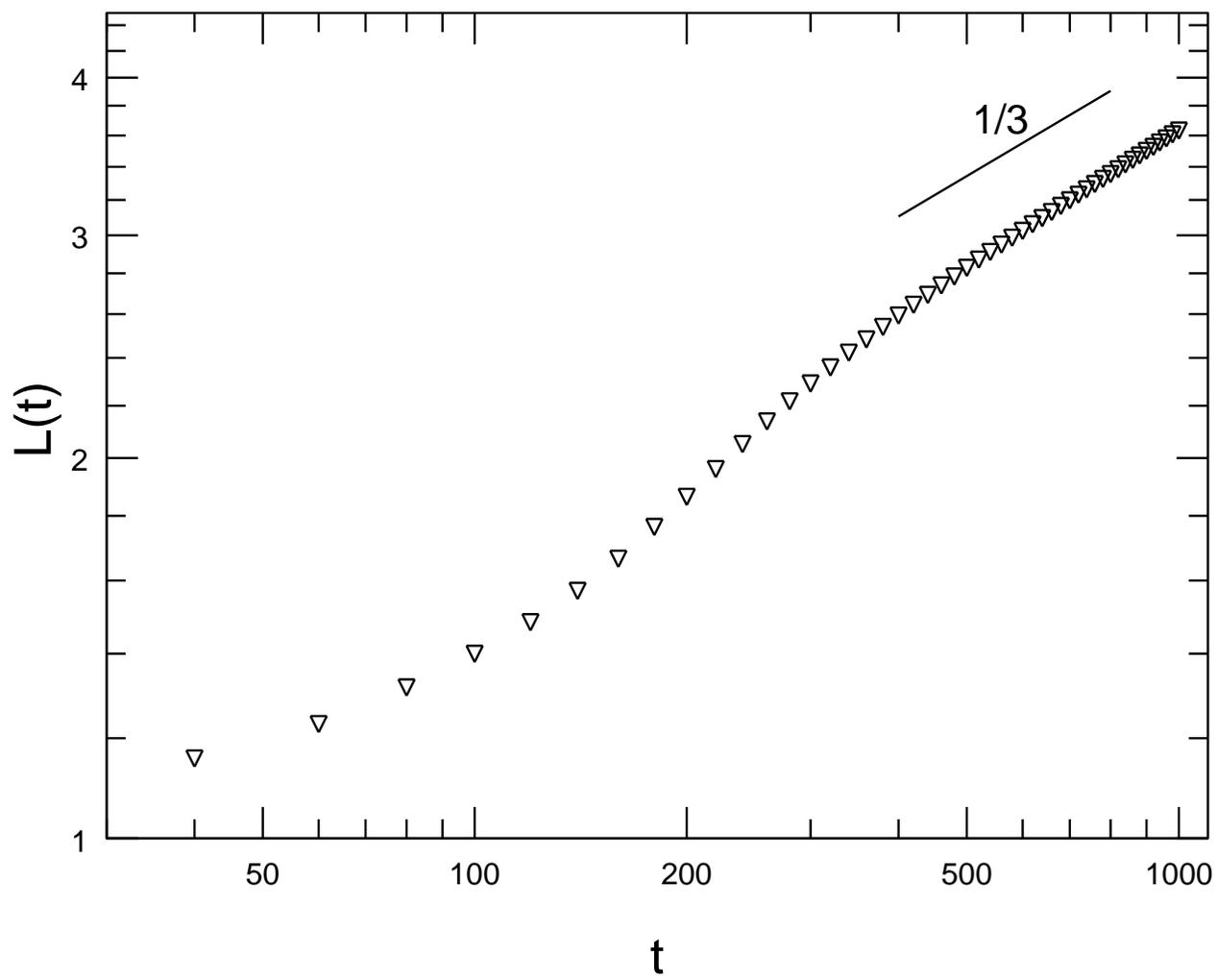
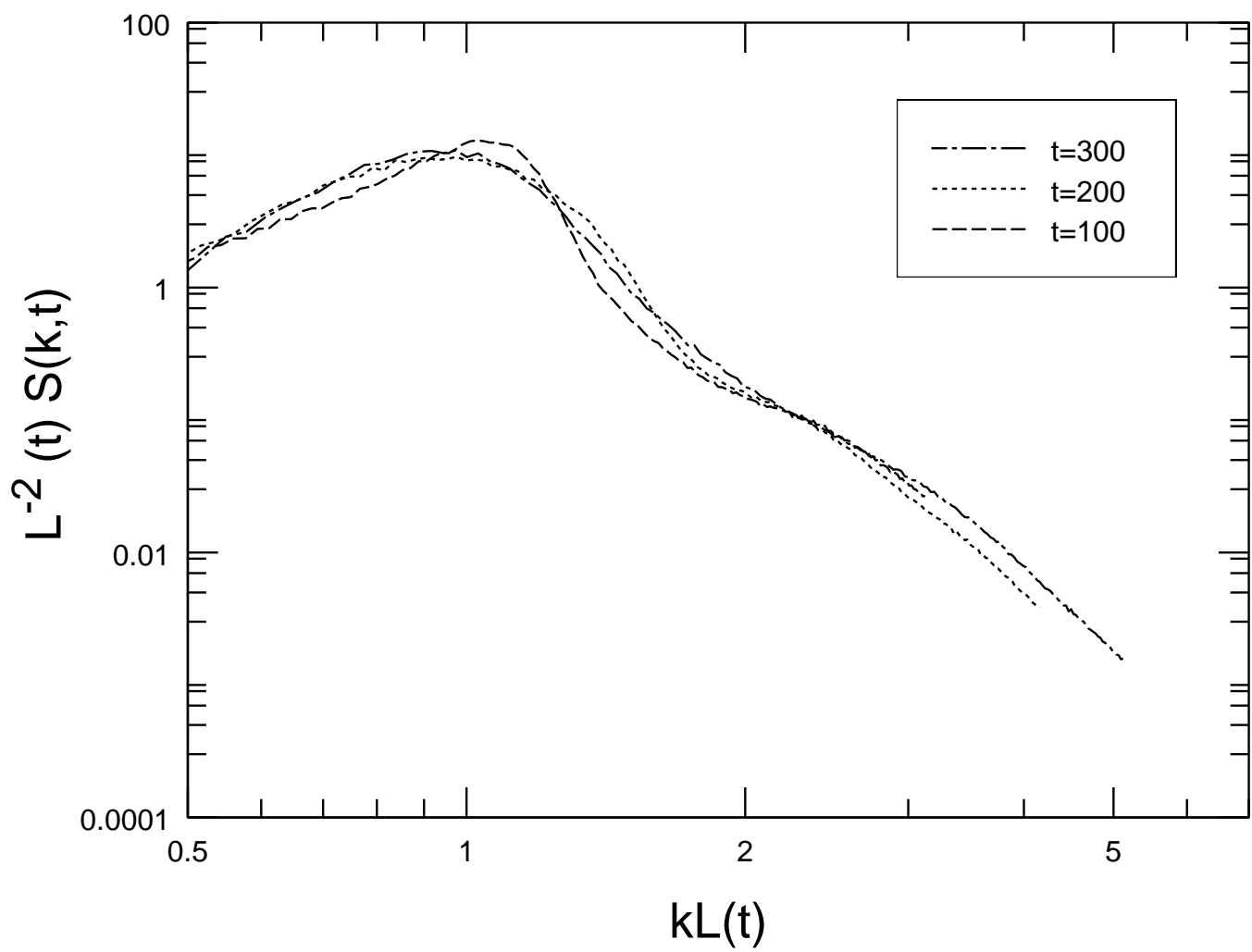
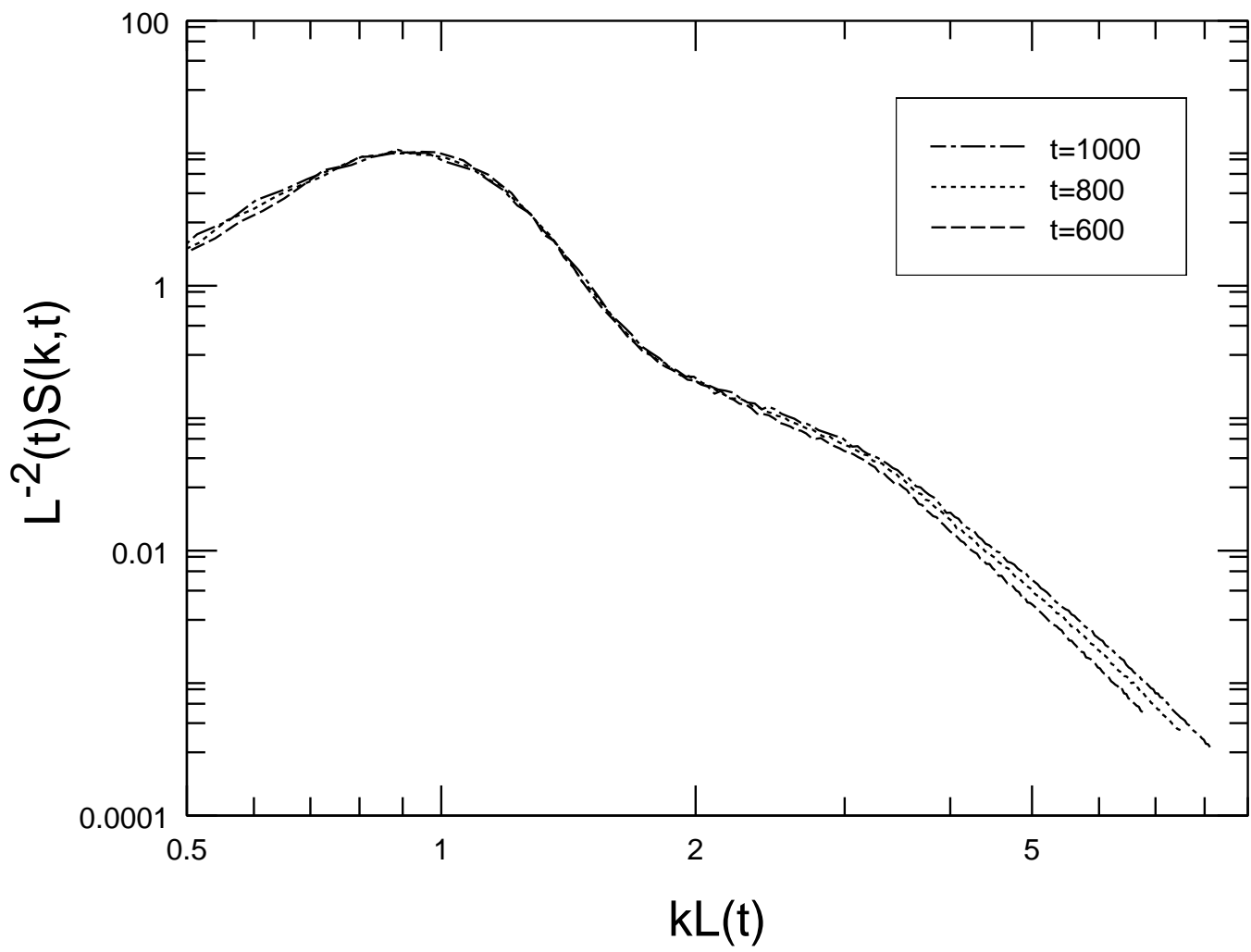
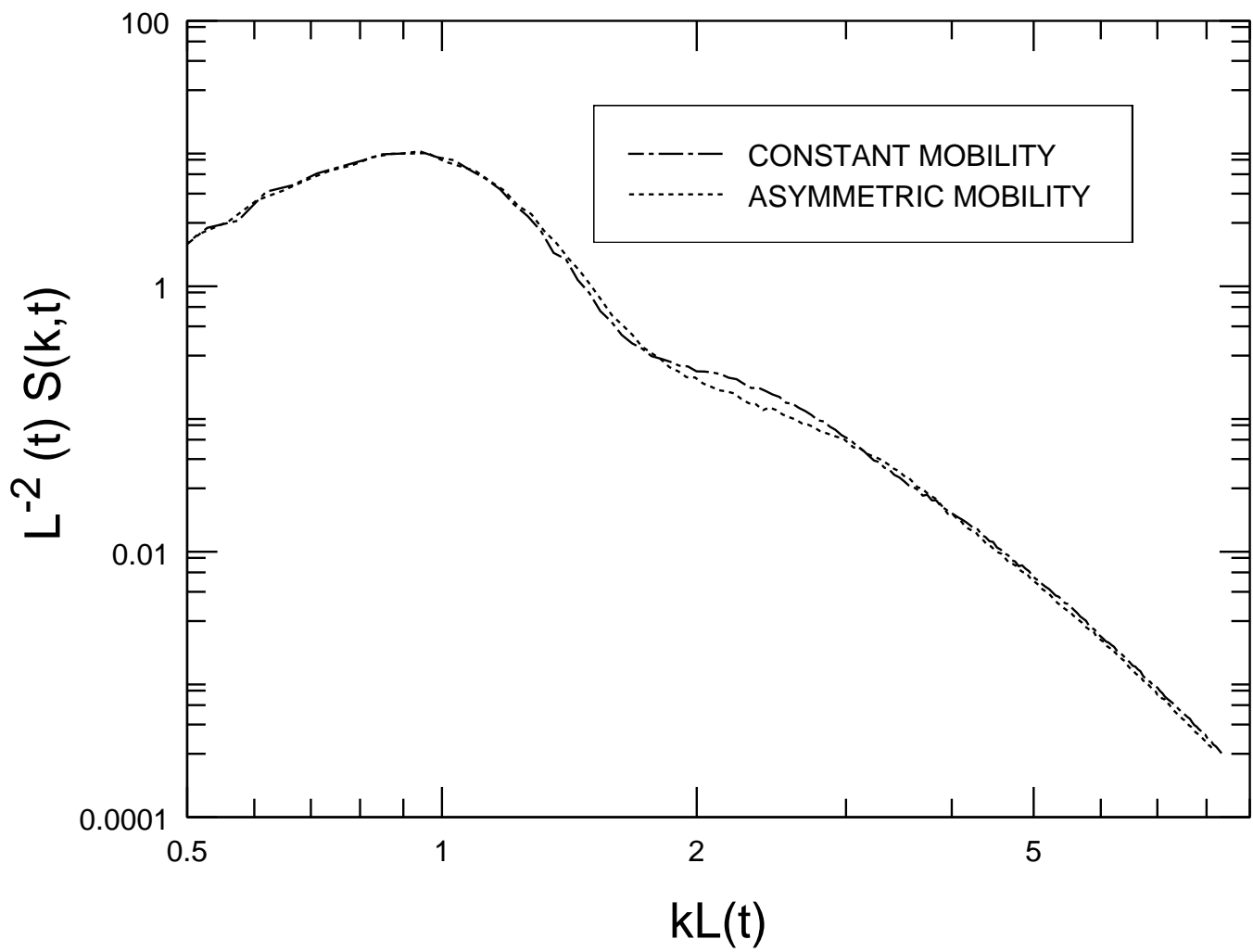
\end{document}